\documentclass[journal,twocolumn]{IEEEtran}
\usepackage[numbers,sort&compress]{natbib}
\usepackage{graphicx}
\usepackage{graphics}
\usepackage{latexsym}
\usepackage{amssymb}
\usepackage{verbatim}
\usepackage{amsmath}
\usepackage{amsthm}
\usepackage{amsfonts}

\newtheorem{Theorem}{Theorem}
\newtheorem{Proposition}[Theorem]{Proposition}
\newtheorem{Definition}[Theorem]{Definition}

\newtheorem{Example}[Theorem]{Example}
\newtheorem{remark}[Theorem]{Remark}

\ifCLASSINFOpdf
\else
\fi
\usepackage[tight,footnotesize]{subfigure}

\usepackage{stfloats}

\begin{document}
%
\title{Turbo Lattices: Construction and Error Decoding Performance}
%
%
%

\author{Amin~Sakzad,~\IEEEmembership{Member,~IEEE,}
        Mohammad-Reza~Sadeghi, 
        and~Daniel~Panario,~\IEEEmembership{Senior~Member,~IEEE}
\thanks{A. Sakzad and M. R. Sadeghi are with the
Department of Mathematics and Computer Science, Amirkabir University of Technology,
Emails: amin$\_$sakzad@aut.ac.ir and msadeghi@aut.ac.ir.}
\thanks{D. Panario is with the School of Mathematics and Statistics, Carleton University,
Email: daniel@math.carleton.ca.}
\thanks{This paper was presented in part at the 48th Annual Allerton Conference on Communication, Control, and Computing, Allerton 2010.}}

\maketitle

\begin{abstract}
In this paper a new class of lattices called turbo lattices
is introduced and established.
We use the lattice Construction D to produce turbo lattices.
This method needs a set of nested linear codes as its underlying
structure. We benefit from turbo codes as our basis codes.
Therefore, a set of nested turbo codes based on
nested interleavers (block interleavers) and nested convolutional codes is built.
To this end, we employ both tail-biting and zero-tail convolutional codes.
Using these codes, along with construction D, turbo lattices
are created. Several properties of Construction D
lattices and fundamental characteristics of turbo lattices
including the minimum distance, coding gain and kissing number
are investigated. Furthermore, a multi-stage
turbo lattice decoding algorithm based on iterative turbo decoding algorithm
is given.
We show, by simulation, that turbo lattices attain
good error performance within $\sim1.25~\mbox{dB}$ from capacity at block length of
$n=1035$. Also an excellent performance of only $\sim.5~\mbox{dB}$ away from capacity
at SER of $10^{-5}$ is achieved for size $n=10131$.

\end{abstract}


%
\IEEEpeerreviewmaketitle

\section{Introduction}~\label{Introduction}
%
%
%
%
Turbo codes were first introduced by Berrou {\em et al.}~\cite{berrou} in 1993 and
have been largely treated since then. It has been shown~\cite{costello} that these codes with an
iterative turbo decoding algorithm can achieve a very good error performance close
to Shannon capacity. Also, there has been interest in constructing lattices with high coding gain,
low kissing number and low decoding complexity~\cite{IrregularLDPC, sadeghi, LDLC}.
The lattice version of the channel coding is
to find an $n$-dimensional lattice $\Lambda$ which attains good error performance
for a given value of volume-to-noise ratio (VNR)~\cite{conway, forneyspherebound, tarokh}.
Poltyrev~\cite{polytrev} suggests employing coding without restriction
for lattices on the AWGN channel. This means communicating
with no power constraints.
The existence of ensembles of lattices which can achieve generalized capacity
on the AWGN channel without restriction is also proved in~\cite{polytrev}.
Forney {\em et al.}~\cite{forneyspherebound}
restate the above concepts by using coset codes and multilevel
coset codes. At the receiver of communication without restriction for lattices,
the main problem is to find the closest
vector of $\Lambda$ to a given point ${\bf r}\in\mathbb{R}^n$.
This is called lattice decoding of $\Lambda$.
Some efficient well-known lattice decoders are known for low dimensions~\cite{Ling09,Viterbo99}.

There are a wide range of applicable lattices in communications
including the well-known root lattices~\cite{conway}, the recently
introduced low-density parity-check lattices~\cite{sadeghi} (LDPC
lattices) and the low-density lattice codes~\cite{LDLC} (LDLC lattices).
The former lattices have been extensively treated in the 1980's
and 1990's~\cite{conway}. After the year 2000, two classes of
lattices based on the primary idea of LDPC codes have been
established. These type of lattices have attracted a lot of
attention in recent years~\cite{IrregularLDPC,choi,sadeghi2,Harshan13}.
Hence, constructing lattices based on turbo codes can be a promising research topic.

In the present work, we borrow the idea of turbo codes and construct a new class of
lattices that we called {\em turbo lattices}. In fact, the results by Forney {\em et al.} in~\cite{forneyspherebound}
motivate us to apply Construction D lattices to design turbo lattices.
They proved the existence of sphere-bound-achieving lattices by means of Construction D
lattices. This leads one to use Construction D method along
with well-known turbo codes to produce turbo lattices.
This is the first usage of turbo codes in constructing lattices.
We benefit from structural properties of lattices and turbo
codes to investigate and evaluate the basic parameters of turbo
lattices such as minimum distance, volume, coding gain and
kissing number.

Various types of turbo codes have been constructed in terms of properties of their underlying
constituent encoders and interleavers~\cite{costello}. For example,
encoders can be either block or convolutional codes and
interleavers can be deterministic, pseudo-random or random~\cite{recent}.
Since Construction D deals with block codes, we treat turbo codes as block codes.
Therefore, it seems more reasonable to use terminated convolutional
codes. Since we use recursive and non-recursive convolutional codes,
different types of termination methods can be applied to these component convolutional codes.
Hence, we are interested in terminating trellises for
both feed-back~\cite{solomon,wiss} and feed-forward~\cite{costello}
convolutional codes. To stay away from rate loss, we employ tail-biting
convolutional codes for short length turbo lattices. Also zero-tail
convolutional codes~\cite{solomon,wiss} are building blocks of turbo codes to use in construction of
lattices with larger sizes.

There are algorithms such as generalized min-sum algorithm~\cite{sadeghi},
iterative decoding algorithms~\cite{choi} and the algorithm in~\cite{LDLC} for decoding
newly introduced lattices.
The basic idea behind these algorithms
is to implement min-sum and sum-product algorithms and their generalizations.
Since we used turbo codes to construct
turbo lattices, it is more reasonable to benefit
from the underlying turbo structure of these lattices.
In this case, we have to somehow relate the decoding of turbo lattices
to the iterative turbo decoders~\cite{berrou} for turbo codes.
This results in a multi-stage decoding algorithm based on
iterative turbo decoders similar to the one given in~\cite{forneyspherebound}.

We summarize our contributions as follows.
\begin{itemize}
\item We generalize the minimum distance formula for every
Construction D lattice by removing a restricting condition on the minimum distance of its underlying codes.
An upper bound for the kissing number of these lattices is also derived.
\item We construct nested turbo codes and establish the concept of turbo lattices. Various crucial parameters of
these lattices such as minimum distance, coding gain and kissing number are investigated.
\item A multi-stage turbo lattice decoder is introduced. The error performance of turbo lattices
is given and compared with other well-known LDPC lattices and LDLC lattices.
\end{itemize}

The present work is organized as follow. Two methods of constructing
lattices, Construction A and D, are reviewed in Section~\ref{BackgroundsonLattices}.
The crucial parameters of lattices which can be used to measure the
efficiency of lattices are explained in this section. In Section~\ref{ConvolutionalandTurboCodes} we
introduce nested interleavers in a manner that can be used to build nested turbo codes.
Section~\ref{NestedTCandTL} is devoted to the construction of nested turbo codes and consequently
the construction of turbo lattices.
Section~\ref{ParameterAnalysis} is dedicated to the evaluation
of the critical parameters of turbo lattices
based on the properties of their underlying turbo codes. In Section~\ref{DecodingAlgorithm}
a multi-stage turbo lattice decoding algorithm is explained.
In Section~\ref{SimulationResultsPerformanceofTL}
we carry simulation results.
We conclude with final remarks on turbo lattices and further
research topics in Section~\ref{Conclusion}.

\section{Backgrounds on Lattices}~\label{BackgroundsonLattices}
In order to make this work self-contained, a background on
lattices is essential. The general required information about
critical parameters of Construction A and Construction D
as well as parameters for measuring the efficiency of lattices
are provided below.
\subsection{General Notations for Lattices}~\label{lattices}
A discrete additive subgroup $\Lambda$ of $\mathbb{R}^n$
is called \emph{lattice}. Since $\Lambda$ is discrete, it can be generated by
$m\leq n$ linearly independent vectors ${\bf b}_1,\ldots,{\bf b}_m$ in
$\mathbb{R}^n$. The set $\{{\bf b}_1,\ldots,{\bf b}_m\}$ is called a \emph{basis}
for $\Lambda$.
In the rest of this paper, we assume that $\Lambda$ is an
$n$-dimensional full rank ($m=n$) lattice over $\mathbb{R}^n$.
By using the Euclidean norm, $\|.\|$, we can define a metric on
$\Lambda$; that is, for every ${\bf x},{\bf y}\in\Lambda$ we have $d({\bf x},{\bf y})=\|{\bf x}-{\bf y}\|^2$.
The \emph{minimum distance} of $\Lambda$, $d_{\min}(\Lambda)$,
is
$$d_{\min}(\Lambda)=\min_{{\bf x}\neq{\bf y}}\{d({\bf x},{\bf y})|{\bf x},{\bf y}\in \Lambda\}.$$
Let us put $\{{\bf b}_1,\ldots,{\bf b}_n\}$ as the rows of a matrix ${\bf B}$, then
we have $\Lambda=\{{\bf x}\colon {\bf x}={\bf z}{\bf B},~{\bf z}\in \mathbb{Z}^n\}$.
The matrix ${\bf B}$ is called a \emph{generator matrix} for the lattice $\Lambda$. The \emph{volume}
of a lattice $\Lambda$ can be defined by $\det\left({\bf B}{\bf B}^T\right)$ where
${\bf B}^T$ is the transpose of ${\bf B}$. The volume of $\Lambda$
is denoted by $\det(\Lambda)$.
The {\em coding gain} of
a lattice $\Lambda$ is defined by
\begin{equation}~\label{eq:codinggain}
\gamma(\Lambda)=\frac{d_{\min}^2(\Lambda)}{\det(\Lambda)^{2/n}},
\end{equation}
where $\det(\Lambda)^{2/n}$ is itself called the
\emph{normalized volume} of $\Lambda$.
This volume may be regarded as the volume of $\Lambda$ per two dimensions.
The coding gain can be used as a crude measure of the performance of a lattice.
For any $n$, $\gamma\left(\mathbb{Z}^n\right)=1$. An uncoded system
may be regarded as the one that uses a constellation based on $\mathbb{Z}^n$.
Thus the coding gain of an arbitrary lattice $\Lambda$ may be considered
as the gain using a constellation based on $\Lambda$ over an uncoded system
using a constellation based on $\mathbb{Z}^n$~\cite{forneyspherebound}. 
Therefore, coding gain is the saving in average of energy due to using $\Lambda$
for the transmission instead of using the lattice $\mathbb{Z}^n$~\cite{forneymodulation}.
Geometrically, coding gain measures the increase
in density of $\Lambda$ over integer lattice $\mathbb{Z}^n$~\cite{conway}.

If one put an $n$-dimensional sphere of
radius $d_{\min}(\Lambda)/2$ centered at
every lattice point of $\Lambda$,
then the {\em kissing number} of $\Lambda$
is the maximum number
of spheres that touch a fixed sphere.
Hereafter we denote the kissing number of the lattice $\Lambda$
by $\tau(\Lambda)$. The \emph{normalized kissing number} of an $n$-dimensional
lattice $\Lambda$ is defined as
\begin{equation}\label{eq:normalizedkissnum1}
 \tau^\ast(\Lambda)=\frac{\tau(\Lambda)}{n}.
\end{equation}

Sending points of a specific lattice in the absence of power constraints has been
studied. This is called {\em coding without restriction}~\cite{polytrev}.
Suppose that the points of an $n$-dimensional lattice $\Lambda$ are
sent over an AWGN channel with noise variance $\sigma^2$.
The {\it volume-to-noise ratio} (VNR) of an $n$-dimensional lattice $\Lambda$ is defined as
\begin{equation}~\label{VNR}
{\alpha^2}=\frac{\det(\Lambda)^{\frac{2}{n}}}{2\pi e\sigma^2}.
\end{equation}
For large $n$, the VNR is the ratio of the normalized volume of $\Lambda$
to the normalized volume of a noise sphere of squared radius $n\sigma^2$
which is defined as SNR in~\cite{sadeghi} and $\alpha^2$ in~\cite{forneyspherebound}.

Since lattices have a uniform structure,
we can assume ${\bf 0}$ is transmitted and $\textbf{r}$
is the received vector. Then $\textbf{r}$ is a vector whose components are
distributed based on a Gaussian distribution with zero mean and variance $\sigma^2$.
Hence construction of lattices with higher coding gain and lower normalized
kissing number is of interest.

\subsection{Lattice Constructions}~\label{Construction A}
There exist many ways to construct a
lattice~\cite{conway}.
In the following we give two algebraic constructions of lattices based
on linear block codes~\cite{conway}. The first one is Construction A
which translates a block code to a lattice.
Then a review of Construction D is given.
These two constructions are the main building blocks of this work.

Let $\mathcal{C}$ be a group code over
$G=\mathbb{Z}_{2}\times\cdots\times \mathbb{Z}_{2}$
, i.e. $\mathcal{C}\subseteq G$, with minimum distance $d_{\min}$. Define
$\Lambda$ as a Construction A lattice~\cite{conway} derived from $\mathcal{C}$ by:
\begin{equation}\label{constA}
\Lambda=\{(2z_1+c_1,\ldots,2z_n+c_n): z_i\in\mathbb{Z}, {\bf c}=(c_1,\ldots,c_n)\in \mathcal{C}\}.
\end{equation}
Let $\Lambda$ be a lattice constructed using Construction A.
The minimum distance of $\Lambda$ is
\begin{equation}~\label{eq:dminconstA}
d_{\min}(\Lambda)=\min\left\{2,\sqrt{d_{min}}\right\}.
\end{equation}
Its coding gain is
\begin{equation}~\label{eq:codinggainconstA}
\gamma(\Lambda)=
\left\{\begin{array}{ll}
4^{\frac{k}{n}}&d_{\min}\geq4,\\
\frac{d_{\min}^2(\Lambda)}{2}4^{\frac{k}{n}}&d_{\min}<4,
\end{array}\right.
\end{equation}
and its kissing number is
\begin{equation}~\label{eq:kissnumberconstA}
\tau(\Lambda)=
\left\{\begin{array}{ll}
2^{d_{\min}}A_{d_{\min}}&d_{\min}<4,\\
2n+16A_4&d_{\min}=4,\\
2n&d_{\min}>4,
\end{array}\right.
\end{equation}
where $A_{d_{\min}}$ denotes the number of codewords in
$\mathcal{C}$ with minimum weight $d_{\min}$.
These definition and theorem can be generalized to a more
practical and nice lattice construction.
We use a set of nested linear block codes to
give a more general lattice structure named
Construction D. This construction plays a key role
in this work.

Let $\mathcal{C}_0\supseteq \mathcal{C}_1\supseteq \cdots \supseteq \mathcal{C}_a$
be a family of $a+1$ linear codes where $\mathcal{C}_{\ell}[n,k_{\ell},d_{\min}^{(\ell)}]$
for $1\leq {\ell}\leq a$ and
$\mathcal{C}_0$ is the $[n,n,1]$ trivial code $\mathbb{F}_2^n$ such that
$$\mathcal{C}_{\ell}=<{\bf c}_1,\ldots,{\bf c}_{k_{\ell}}>$$
where $<X>$ denotes the subgroup generated by $X$.
For any element ${\bf x}=(x_1,\ldots,x_n)\in \mathbb{F}^n_2$ and for
$1\leq {\ell}\leq a$ consider the vector in $\mathbb{R}^n$ of the form:
$$\frac{1}{2^{{\ell}-1}}{\bf x}=\left(\frac{x_1}{2^{{\ell}-1}},\ldots,\frac{x_n}{2^{{\ell}-1}}\right).$$
Define $\Lambda\subseteq\mathbb{R}^n$
as all vectors of the form
\begin{equation}~\label{eq:formofconstructionD}
{\bf z}+\sum_{{\ell}=1}^a\sum_{j=1}^{k_{\ell}}\beta_j^{({\ell})}\frac{1}{2^{{\ell}-1}}{\bf c}_j
\end{equation}
where ${\bf z}\in2(\mathbb{Z})^n$ and $\beta_j^{({\ell})}=0$ or $1$.
An integral
basis for $\Lambda$ is given by the vectors
\begin{equation}~\label{eq:integralbasisforD}
\frac{1}{2^{{\ell}-1}}{\bf c}_j
\end{equation}
for $1\leq {\ell}\leq a$ and $k_{{\ell}+1}+1\leq j\leq k_{\ell}$ plus $n-k_1$
vectors of the form $(0,\ldots,0,2,0,\ldots,0)$. Let us consider vectors
${\bf c}_j$ as integral in $\mathbb{R}^n$, with components $0$ or $1$.
To be specific,  this lattice $\Lambda$ can be represented by the following code formula
\begin{equation}~\label{eq:codeformulaforD}
\Lambda=\mathcal{C}_1+\frac{1}{2}\mathcal{C}_2+\cdots+\frac{1}{2^{a-1}}\mathcal{C}_{a}+2(\mathbb{Z})^n.
\end{equation}

It is useful to bound the coding gain of $\Lambda$.
The next theorem is cited form~\cite{barnes}.
\begin{Theorem}
Let $\Lambda$ be a lattice constructed using
Construction D, then the volume of $\Lambda$ is
$\det(\Lambda)=2^{n-\sum_{{\ell}=1}^a k_{\ell}}$.
Furthermore, if $d_{\min}^{(\ell)}\geq\frac{4^{\ell}}{\beta}$,
for $1\leq {\ell}\leq a$ and $\beta=1$ or $2$,
then the squared minimum distance of $\Lambda$ is at
least $4/\beta$,
and its coding gain satisfies
$$\gamma(\Lambda)\geq\beta^{-1}4^{\sum_{{\ell}=1}^a\frac{k_{\ell}}{n}}.$$
\end{Theorem}
In the above theorem, an exact formula for the determinant of
every lattice constructed using Construction
D is given. Also, proper bounds for
the other important parameters of these lattices including
minimum distance and coding gain have been found with an extra condition on the
minimum distance of the underlying nested codes~\cite{conway}.

We omit this restricting condition on the minimum distance
of the underlying nested block codes and then generalize those bounds
to a more useful form. The resulting expressions for minimum distance
and coding gain are related to the underlying codes
as we will see soon. In addition, an upper bound for the kissing number
of every lattice generated using Construction D is derived.
\begin{Theorem}~\label{th:constructionDmindistkiss}
Let $\Lambda$ be a lattice constructed based
on Construction D. Then
\begin{itemize}
\item{}
for the minimum distance of $\Lambda$ we have
\begin{equation}~\label{eq:mindistD}
d_{\min}(\Lambda)=\min_{1\leq {\ell}\leq a}\left\{2,\frac{\sqrt{d_{\min}^{(\ell)}}}{2^{{\ell}-1}}\right\},
\end{equation}
where $d_{\min}^{(\ell)}$ is the minimum distance of $\mathcal{C}_{\ell}$ for $1\leq {\ell}\leq a$;
\item{}the kissing number of $\Lambda$ has the following property
\begin{equation}~\label{eq:ubkiss}
\tau(\Lambda)\leq2n+\sum_{\substack{1\leq {\ell}\leq a\\ d_{\min}^{(\ell)}\leq4^{\ell}}}2^{d_{\min}^{(\ell)}}A_{d_{\min}^{(\ell)}},
\end{equation}
where $A_{d_{\min}^{(\ell)}}$ denotes the number of codewords
in $\mathcal{C}_{\ell}$ with minimum weight $d_{\min}^{(\ell)}$. Furthermore, if $d_{\min}^{(\ell)}>4^{\ell}$ for every $1\leq \ell\leq a$,
then $\tau(\Lambda)\leq 2n$.
\end{itemize}
\end{Theorem}
The proof is given in Appendix A.

This theorem provides a relationship between
the performance of the lattice $\Lambda$
and the performance of its underlying codes.
The kissing number of a Construction D lattice can be bounded
above based on the minimum distance and the number of minimum weight
codewords of each underlying nested code.
\section{Convolutional and Turbo Codes}~\label{ConvolutionalandTurboCodes}
Since recursive convolutional codes produce better turbo codes,
we focus on tail-biting of feed-back convolutional codes.
\subsection{Terminated and Tail-Biting Convolutional Codes}~\label{TerminatedConvolutionalCodes}
Let $\mathcal{C}(N,K,\nu)$ be a systematic convolutional code of rate $\frac{K}{N}$
with constraint length $\nu$ and memory order $m$.
The {\em terminated convolutional code} technique can be found in~\cite{costello}.
It is known that, in this deformation from the convolutional code $\mathcal{C}$
to the mentioned block code there exists a rate loss and
a change in the size of the codewords
while in Construction D
all the code lengths of the set of nested linear codes
have to be equal. However, this termination method modifies the sizes of the underlying
codes in each level. This code length modification
results in a restriction which prevents the use of
terminated convolutional codes in our derivation of lattices based on Construction D.
In order to avoid this situation, an alternative method
which is referred as tail-biting~\cite{solomon} can be used. Thus, terminated convolutional codes
can only be employed to construct turbo codes which are appropriate for using along with
Construction A.

The {\em tail-biting} technique for feed-forward convolutional codes are reported in~\cite{conferenceAllerton,solomon,wiss}.
The algorithm for tail-biting a feed-back convolutional encoder
is also introduced in~\cite{theory,wiss}.
However, tail-biting is impossible for all sizes.
In other words, tail-biting of a feed-back convolutional encoder
is only possible for some special tail-biting lengths.

Let ${\bf G}(x)$ be a generator matrix of a systematic feed-back
convolutional code $\mathcal{C}(N,K,\nu)$ defined as follows
\begin{equation}\label{eq:genmatrixSRCC}
\left[
\begin{array}{cccccc}
1&\cdots&0&g_{1,K+1}(x)&\cdots&g_{1,N}(x)\\
\vdots&\ddots&\vdots&\vdots&\ddots&\vdots\\
0&\cdots&1&g_{K,K+1}(x)&\cdots&g_{K,N}(x)
\end{array}
\right],
\end{equation}
where $g_{i,j}(x)=\frac{q_{i,j}(x)}{r_i(x)}$ for coprime
polynomials $q_{i,j}(x)$ and $r_i(x)$, $1\leq i\leq K$
and $K+1\leq j\leq N$.
By means of tail-biting~\cite{conferenceAllerton}, we can corresponds a rate $\frac{K}{N}$ systematic feed-back
convolutional encoder with constraint $\nu$ and a linear code $[LN,LK]$
(where $L$ is called {\em tail-biting length})
with generator matrix
\begin{equation}\label{eq:genmatrixtailbited}
{\bf G}'=\left[
\begin{array}{ccccccc}
{\bf R}_1&\cdots&{\bf 0}&{\bf Q}_{1,K+1}&\cdots&{\bf Q}_{1,N}\\
\vdots&\ddots&\vdots&\vdots&\ddots&\vdots\\
{\bf 0}&\cdots&{\bf R}_K&{\bf Q}_{K,K+1}&\cdots&{\bf Q}_{K,N}
\end{array}
\right],
\end{equation}
where ${\bf Q}_i$ and ${\bf Q}_{i,j}$ are $L\times L$ circulant
matrices with top row of length $L$ made from $r_i(x)$ and $q_{i,j}(x)$ respectively
for $1\leq i\leq K$ and $K+1\leq j\leq N$.
\begin{Theorem}~\label{th:converCCsys}
Let $r_i(x),~q_{i,j}(x),~L$ and ${\bf R}_i, {\bf Q}_{i,j}$
be as above for $1\leq i\leq K$ and $K+1\leq j\leq N$.
Then the block code $\mathcal{C}[LN,LK]$ generated by ${\bf G}'$ in~(\ref{eq:genmatrixtailbited})
can also be generated by
${\bf G}=[{\bf I}_{LK}|{\bf F}]$, where ${\bf F}$ is a circulant matrix if and only if
$(r_i(x),x^L-1)=1$ for all $1\leq i\leq K$. In this case,
we get
$$q_{i,j}(x)\equiv f_{i,j}(x)r_i(x)\!\!\pmod{x^L-1}.$$
\end{Theorem}
The proof is given in Appendix A.

We observe that ${\bf F}$ is an $LK\times L(N-K)$ circulant
matrix consisting of $K\times (N-K)$ blocks of $L\times L$
circulant submatrices  which must be placed
in the $(i,j)$--th block of ${\bf F}$. It is obtained using $f_{i,j}(x)$
as its top row, $1\leq i\leq K$ and $K+1\leq j\leq N$.
Also the identity matrix ${\bf I}_{LK}$ can be written as an $K\times K$ identity block matrix
with each of its nonzero entries replaced by an identity matrix ${\bf I}_{L}$.

We close this subsection giving a proposition that relates our result in the above theorem
and well-known results~\cite{sathl,wiss} for eligible lengths of $L$
that can be applied to construct tail-biting feed-back
convolutional codes. For the sake of brevity, we consider
only feed-back convolutional codes of rate $\frac{N-1}{N}$.
Let ${\bf G}(x)$ be a generator matrix of a systematic feed-back
convolutional code $\mathcal{C}(N,N-1,\nu)$ defined as follows
\begin{equation}\label{eq:genmatrixSRCCforequivalence}
\left[
\begin{array}{cccc}
1&\cdots&0&g_{1,N}(x)\\
\vdots&\ddots&\vdots&\vdots\\
0&\cdots&1&g_{N-1,N}(x)
\end{array}
\right],
\end{equation}
where $g_{i,N}(x)=\frac{q_{i,N}(x)}{r(x)}$ for coprime
polynomials $q_{i,N}(x)$ and $r(x)$ for $1\leq i\leq N-1$.
Without loss of generality, we assume that
$r(x)=r_0+r_1x+\cdots+r_mx^m$.
If we realize this code in observer canonical form~\cite{wiss},
then the state matrix is
\begin{equation}\label{eq:statematrixSRCCforequivalence}
{\bf A}=\left[
\begin{array}{ccc|c}
0&\cdots&0&r_m\\
1&\cdots&0&r_{m-1}\\
\vdots&\ddots&\vdots&\vdots\\
0&\cdots&1&r_1
\end{array}
\right].
\end{equation}
We have that in order to encode an $[LN,LK]$ tail-biting code
with the method described in~\cite{wiss}, the matrix $\left({\bf A}^L-{\bf I}_m\right)$
has to be invertible. It should be noted that~\cite{wiss} realizing the encoder
in controller canonical form and observer canonical form leads to
the same set of possible sizes $L$.
\begin{Proposition}~\label{relation}
Let ${\bf A}$, as in~(\ref{eq:statematrixSRCCforequivalence}),
be the state matrix of a convolutional code $\mathcal{C}(N,N-1,\nu)$
with generator matrix~(\ref{eq:genmatrixSRCCforequivalence}).
Then  $\det\left({\bf A}^L-{\bf I}_m\right)\neq0$ if and only if $\gcd(r(x),x^L-1)=1$.
\end{Proposition}
The proof is given in Appendix A.
\subsection{Parallel Concatenated Codes; Structure of Turbo Codes}~\label{ParallelConcatenatedCodes;StructureofTurboCodes}
Turbo codes can be assumed as block codes by fixing their interleaver lengths;
but they have not been analyzed from this point of view except in~\cite{recent}.
We follow the construction of turbo codes from~\cite{berrou,costello} and then we use them
to produce a new type of lattices called {\em turbo lattices}.
We assume that an interleaver
$\Pi$ and a recursive convolutional encoder $\mathcal{E}$ with parameters
$(N,K,\nu)$ are used for constructing a turbo code of size $k=KL$.

The information block (interleaver size) $k$ has to be selected
large enough to achieve performance close to Shannon limit.
Improving minimum free distance of turbo codes is possible by designing
good interleavers. In other words, interleavers make a shift from lower-weight
codewords to higher-weight codewords. This shifting has been called
{\it spectral thining}~\cite{costello}. Such interleaving matches the
codewords with lower weight of the first encoder to the high-weight
parity sequences of the second encoder. More precisely, for large
values of interleaver size $k$ the multiplicities of
the low-weight codewords in the turbo code weight spectrum are
reduced by a factor of $k$. This reduction by a factor of
$k$ is called \emph{interleaver gain}. Hence, it is apparent that
interleavers have a key role in the heart of turbo codes and it is important to
have random-like properties for interleavers~\cite{costello,recent}.
Boutros et. al provided almost optimal interleavers in~\cite{Boutros06}.
\section{Nested Turbo Codes and Turbo Lattices}~\label{NestedTCandTL}
We exploit a set of nested tail-biting convolutional codes and a nested interleaver
along with Construction D to form turbo lattices.
Also terminated convolutional codes and Construction A
are employed for the same purpose. An explicit explanation of these two approaches is given next.
\subsection{Constructing Nested Turbo Codes}~\label{ConstructingNestedTurboCodes}
Consider a turbo code $\mathcal{TC}$ with two component codes generated
by a generator matrix ${\bf G}(x)$ of size ${K\times N}$ of a convolutional code
and a random interleaver $\Pi$, of size $k=LK$.
Assume that both encoders are systematic feed-back convolutional codes.
Every interleaver $\Pi$ can be represented by a matrix ${\bf P}_{k\times k}$
which has only a single $1$ in each column and row. It is easy to see that
the generator matrix of $\mathcal{TC}$ can be written as follows
\begin{equation}~\label{eq:genmatTC}
{\bf G}_{\mathcal{TC}}=
\left[\begin{array}{c|c|c}
{\bf I}_k&{\bf F}&{\bf PF}
\end{array}\right]_{k\times n}
\end{equation}
where ${\bf F}$ is a $LK\times L(N-K)$ submatrix of ${\bf G}$, the tail-bited generator matrix
of ${\bf G}(x)$, including only parity columns of ${\bf G}$. The matrix ${\bf I}_k$ is the identity matrix of size $k$.
Therefore, we can assume that ${\bf G}_{\mathcal{TC}}$
is a $k\times n$ matrix with $k=LK$ rows and $n=2LN-LK$ columns.

The above representation~(\ref{eq:genmatTC}) can be extended to construct a generator matrix
for a parallel concatenated code with $b$ branches. Each branch has its
own interleaver $\Pi_j$ with matrix representation
${\bf P}_j$ and a recursive encoder $\mathcal{E}_j$ for $1\leq j\leq b$. Assume
that all the encoders are the same $(N, K, \nu)$ convolutional encoder and the block of information bits has length $k=KL$.
Thus, the corresponding generator matrix of this turbo code is
\begin{equation}~\label{eq:genmatextendedTC}
{\bf G}_{\mathcal{TC}}^e=
\left[\begin{array}{c|c|c|c|c}
{\bf I}_k&{\bf P}_1{\bf F}&{\bf P}_2{\bf F}&\cdots&{\bf P}_b{\bf F}
\end{array}\right]_{k\times n_e}
\end{equation}
where ${\bf F}$ is a $LK\times L(N-K)$ as above and $n_e=KL+bL(N-K)$.

In order to design a nested set of turbo codes, the presence of a
nested interleaver is essential. Hence, a new concept
of nested interleavers has to be given.
\begin{Definition}~\label{def:nestedinteleaver}
The interleaver $\Pi$ of size $k$ is a \emph{$(k_a,\ldots,k_1)$-nested
interleaver} if the following conditions hold
\begin{enumerate}
\item{}$0<k_a<k_{a-1}<\cdots<k_1=k$,
\item{}for every $1\leq {\ell}\leq a$, if 
$x\in\{1,\ldots,k_{\ell}\}$, then
$$\Pi(x)\in\{1,\ldots,k_{\ell}\}.$$
\end{enumerate}
A $(k_2,k_1)$-nested interleaver is called a $k_2$-nested interleaver.
\end{Definition}
\begin{Example}~\label{ex:nestedinter}
Let $K=1$. The permutation
$$\Pi_1=
\left(
\begin{array}{ccccccccc}
1&2&3&4&5&6&7&8\\
3&4&2&1&5&7&8&6
\end{array}
\right)$$
is a $4$-nested interleaver because $k_1=k=L=8$ and $k_2=4$.
\end{Example}
The following nested turbo codes
are appropriate to use in both Construction A and Construction D
for producing turbo lattices.
\begin{Definition}~\label{def:nestedTC}
Let $\mathcal{TC}$ be a parallel concatenated convolutional code with
two equivalent systematic convolutional codes generated by ${\bf G}(x)$.
Let ${\bf G}$ be the generator matrix of tail-biting of ${\bf G}(x)$, and
$\Pi$ be the interleaver of size $k=LK$
with the $(k_a,\ldots,k_1)$-nested property
that is used to construct a turbo code $\mathcal{TC}$.
Then ${\bf G}_{\mathcal{TC}}$ is as of~(\ref{eq:genmatTC}).
Define a set of turbo codes
\begin{equation}~\label{eq:nestedturbocodes}
\mathcal{TC}=\mathcal{TC}_1\supseteq \mathcal{TC}_2\supseteq\cdots\supseteq \mathcal{TC}_a.
\end{equation}
In fact, a generator matrix ${\bf G}_{\ell}$ of size
$k_{\ell}\times n$ is a submatrix of ${\bf G}_{\mathcal{TC}}$ consisting of the first
$k_{\ell}$ rows of ${\bf G}_{\mathcal{TC}}$ for every $1\leq {\ell}\leq a$.
\end{Definition}
\begin{Example}~\label{ex:classicaltimedomian}
Consider a $(4,3,3)$ systematic convolutional code
with the following generator matrix
$${\bf G}(x)=\left[
\begin{array}{cccc}
1&0&0&\frac{1+x+x^3+x^4}{1+x^2+x^4}\\
0&1&0&\frac{1+x^3+x^4}{1+x^2+x^4}\\
0&0&1&\frac{1+x+x^2+x^4}{1+x^2+x^4}
\end{array}
\right].$$
The matrix ${\bf G}(x)$ is equivalent to ${\bf G}'(x)$ given by
$$\left[
\begin{array}{cccc}
r_1(x)&0&0&q_{1,4}(x)\\
0&r_2(x)&0&q_{2,4}(x)\\
0&0&r_3(x)&q_{3,4}(x)
\end{array}
\right],$$
where $r_1(x)=r_2(x)=r_3(x)=1+x^2+x^4$ and also
$q_{1,4}(x)=1+x+x^3+x^4$, $q_{2,4}(x)=1+x^3+x^4$ and $q_{3,4}(x)=1+x+x^2+x^4$.
Let $L=8$, then $(r_i(x),x^8-1)=1$ and $q_{i,4}(x)\equiv r_i(x)f_{i,4}(x)\pmod{x^8-1}$, for $1\leq i\leq 3$.
One can use the Euclidean algorithm to find $f_{i,4}(x)$.
Therefore, we get
$$\left\{
\begin{array}{l}
f_{1,4}(x)=x^2+x^3+x^5+x^6,\\
f_{2,4}(x)=x+x^2+x^3+x^6+x^7,\\
f_{3,4}(x)=1+x+x^5+x^7.
\end{array}
\right.$$
Hence,
$${\bf G}=\left[
\begin{array}{cccc}
{\bf I}_8&{\bf 0}&{\bf 0}&{\bf F}_{1,4}\\
{\bf 0}&{\bf I}_8&{\bf 0}&{\bf F}_{2,4}\\
{\bf 0}&{\bf 0}&{\bf I}_8&{\bf F}_{3,4}
\end{array}
\right]_{24\times 32},$$
where ${\bf F}_{i,4}$ is a circulant matrix of size $8$ defined by top row $f_{i,4}(x)$, $1\leq i\leq 3$.
For instance
$${\bf F}_{1,4}=\left[
\begin{array}{cccccccc}
0&0&1&1&0&1&1&0\\
0&0&0&1&1&0&1&1\\
1&0&0&0&1&1&0&1\\
1&1&0&0&0&1&1&0\\
0&1&1&0&0&0&1&1\\
1&0&1&1&0&0&0&1\\
1&1&0&1&1&0&0&0\\
0&1&1&0&1&1&0&0
\end{array}
\right].$$
Assume that ${\bf F}$ is the last $8$ columns of ${\bf G}$, thus
$${\bf F}=\left[
\begin{array}{l}
{\bf F}_{1,4}\\
{\bf F}_{2,4}\\
{\bf F}_{3,4}
\end{array}
\right].$$
Also let us suppose that $\Pi$ is a $(8,16,24)$-nested interleaver
constructed by means of the permutation matrix
$${\bf P}=\left[
\begin{array}{ccc}
{\bf P}_{1,1}&{\bf 0}&{\bf 0}\\
{\bf 0}&{\bf P}_{2,2}&{\bf 0}\\
{\bf 0}&{\bf 0}&{\bf P}_{3,3}\\
\end{array}
\right]_{24\times 24},$$
where ${\bf P}_{j,j}$ is another permutation matrix of size $8\times 8$.
 Then
${\bf G}_{\mathcal{TC}}$, a generator matrix for our nested turbo code is
$${\bf G}_{\mathcal{TC}}=\left[
\begin{array}{c|c|c}
{\bf I}_{24}&{\bf F}&{\bf P}{\bf F}
\end{array}
\right].$$
Now we have $\mathcal{TC}=\mathcal{TC}_1\supseteq \mathcal{TC}_2\supseteq \mathcal{TC}_3$ such that
a generator matrix for $\mathcal{TC}_2$ is consisting of the first $16$ rows and
a generator matrix for $\mathcal{TC}_3$ is consisting of the first $8$ rows
of ${\bf G}_{\mathcal{TC}}$.
\end{Example}

We are prepared to formulate the basic characteristics of nested
turbo codes. Next we study the
structural properties of a set of nested turbo codes in terms
of properties of its subcodes.
Let $\Pi$ be an $(k_a,k_{a-1},\ldots,k_1)$-nested interleaver and
$$\mathcal{TC}=\mathcal{TC}_1\supseteq \mathcal{TC}_2\supseteq\cdots\supseteq \mathcal{TC}_a$$
be a set of nested turbo codes constructed as
above. Then, we have
$d_{\min}=d_{\min}^{(1)}\leq d_{\min}^{(2)}\leq \cdots\leq d_{\min}^{(a)}$
where $d_{\min}^{(\ell)}$ denotes the minimum distance of $\mathcal{TC}_{\ell}$. Also
the rate of $\mathcal{TC}_{\ell}$ is equal to $R_\ell=\frac{k_{\ell}}{n}$
for $1\leq {\ell}\leq a$. Furthermore, we have
$R_{\mathcal{TC}}=R_1\geq R_2\geq \cdots\geq R_a$.
The rate of each $\mathcal{TC}_{\ell}$ can be increased to $\frac{k_{\ell}}{n-k+k_{\ell}}$
because we have $k-k_{\ell}$ all-zero columns in ${\bf G}_{\ell}$.
In fact, these columns can be punctured out to avoid from generating zero bits,
but we can still keep them. Since producing turbo lattices and measuring the performance
of them are in mind, $R_{\ell}=\frac{k_{\ell}}{n}$
is more useful than the actual rate in the turbo lattices.

The upcoming theorem reveals the fact that the rates
of nested turbo codes stay unchanged when
the interleaver sizes are increased. The only impact of this
is on the increasing of the minimum distance (via interleaver gain and spectral thining),
on the coding gain (via change in the numerator not in denominator of the formula)
and on the kissing number of turbo lattices.
These results are shown more explicitly
in Section~\ref{ParameterAnalysis}.
\begin{Theorem}~\label{th:unchengedrate}
Let $\Pi$ be an $(k_a,k_{a-1},\ldots,k_1)$-nested interleaver and
$$\mathcal{TC}=\mathcal{TC}_1\supseteq \mathcal{TC}_2\supseteq\cdots\supseteq \mathcal{TC}_a$$
be a set of nested turbo codes constructed as
above. If we increase
by scaling the tail-biting length $L$ and parameters $k_{\ell}$'s in the construction of the generator matrix of
the turbo codes and induced set of nested turbo codes by a scale factor
of $t$, then the rates of the resulting nested turbo codes remain intact.
\end{Theorem}
The proof is given in Appendix A.

\subsection{Interleaver Design}~\label{InterleaverDesign}
Interleavers play an important role in turbo codes~\cite{Boutros06,sakzad12,recent}. Consequently, a key
point of turbo lattices are also interleavers. They should have random-like
properties and avoid some specific patterns to induce a good minimum distance.
For a turbo code ensemble using uniform interleaver technique, one can
show turbo codes are good in the following sense~\cite{codingtheorem}.
That is, the minimum distance of parallel
concatenated codes with $b$ parallel branches and
recursive component codes grows as $n^{\frac{b-2}{b}}$~\cite{urbankemindist}.
Also the average maximum-likelihood decoder block error probability
approaches zero, at least as fast as $n^{-b+2}$~\cite{codingtheorem}.
Since increase in coding gain and decrease in normalized kissing number is completely
and straightforwardly related to the increase of minimum distance, it is
reasonable to use more than two branches.

We observe that to produce nested turbo codes, an interleaver which satisfies
the $(k_a,\ldots,k_1)$-nested property is necessary. In other words,
we put two conditions in Definition~\ref{def:nestedinteleaver} in a manner that,
along with Definition~\ref{def:nestedTC}, each $\mathcal{TC}_{\ell}$ determines a turbo code.

A method called {\em appending} has been introduced
in order to construct $(k_a,\ldots,k_1)$-nested interleavers
and a detail example for this is provided in~\cite{conferenceAllerton}.
The append operation preserves the deterministic and pseudorandom
properties~\cite{conferenceAllerton}. Indeed, it is clear that if we append $a$ deterministic interleavers,
then a deterministic interleaver can be defined by a function
including at most $a$ cases.

The general picture of a $(k_a,\ldots,k_1)$-nested interleaver
can be viewed as a block interleaver with permutation matrix
$${\bf P}=\left[
\begin{array}{ccc}
{\bf P}_{1,1}&\cdots&{\bf 0}\\
\vdots&\ddots&\vdots\\
{\bf 0}&\cdots&{\bf P}_{a,a}\\
\end{array}
\right]_{k\times k}$$
where ${\bf P}_{\ell,\ell}$ is an $(k_{a+1-\ell}-k_{a+2-\ell})\times(k_{a+1-\ell}-k_{a+2-\ell})$
matrix, $1\leq\ell\leq a+1$ and $k_{a+1}=0$.
The $\ell$--th turbo code $\mathcal{TC}_{\ell}$ is constructed
by the first $k_\ell$ rows of ${\bf P}$.
\subsection{Turbo Lattices}~\label{TL}
Next, turbo codes and their nested versions are used to derive
lattices using Construction D or Construction A.
We use~(\ref{eq:nestedturbocodes}) and their corresponding
generator matrices as nested codes
which we need for producing a lattice based on Construction D.
Now a generator matrix for a lattice constructed using Construction
D can be derived from a set of generator vectors for the largest underlying
code as in~(\ref{eq:integralbasisforD}). Hence, for finding a generator matrix
for $\Lambda$, we have to multiply
the rows of ${\bf G}_{\mathcal{TC}}$ with index numbers between $k_{{\ell}+1}+1$ and $k_{{\ell}}$
by $\frac{1}{2^{{\ell}-1}}$, $0\leq {\ell} \leq a$.
The resulting matrix along with $n-k$ vectors of the form $(0,\ldots,0,2,0,\ldots,0)$
of length $n$ form an integral basis for a lattice $\Lambda_{\mathcal{TC}}$.
\begin{Definition}~\label{def:TL}
A lattice $\Lambda_{\mathcal{TC}}$ constructed using Construction D
is called a turbo lattice if its largest underlying code is a turbo code $\mathcal{TC}$.
\end{Definition}
It is easy to verify that we can form a turbo lattice $\Lambda_{\mathcal{TC}}$
using a turbo code $\mathcal{TC}$ with generator matrix ${\bf G}_{\mathcal{TC}}$ as in~(\ref{eq:genmatTC}).
If the level of our construction, $a$, is larger than $1$, then we have to use
turbo codes which come from tail-bited convolutional codes.
However, if $a=1$ we have a degree of freedom in using a turbo code
built from  either terminated or tail-bited convolutional codes.
\begin{Example}~\label{extailbitingexampleTL}
Let ${\bf G}_{\mathcal{TC}}$ be as in the previous example. In order to obtain a generator matrix
of $\Lambda_{\mathcal{TC}}$, we have multiplied the rows with indices
$\{1,\ldots,8\}$ by $1/4$ and the rows with indices
$\{9,\ldots,16\}$ by $1/2$. The delivered matrix along with
$16$ additional rows of the form $(0,\ldots,0,2,0,\ldots,0)$ produce a generator
matrix for $\Lambda_{\mathcal{TC}}$. Hence, a generator matrix ${\bf G}_{\mathcal{TL}}$ for the
produced turbo lattice is
$${\bf G}_{\mathcal{TL}}=\left[
\begin{array}{rrrrr}
\frac{1}{4}{\bf I}_8&{\bf 0}&{\bf 0}&\frac{1}{4}{\bf F}_{1,4}&\frac{1}{4}{\bf P}_{1,1}{\bf F}_{1,4}\\
{\bf 0}&\frac{1}{2}{\bf I}_8&{\bf 0}&\frac{1}{2}{\bf F}_{2,4}&\frac{1}{4}{\bf P}_{2,2}{\bf F}_{2,4}\\
{\bf 0}&{\bf 0}&{\bf I}_8&{\bf F}_{3,4}&{\bf P}_{3,3}{\bf F}_{3,4}\\
\hline
{\bf 0}&{\bf 0}&{\bf 0}&2{\bf I}_8&{\bf 0}\\
{\bf 0}&{\bf 0}&{\bf 0}&{\bf 0}&2{\bf I}_8
\end{array}
\right]_{40\times 40},$$
where
$${\bf P}=\left[
\begin{array}{ccc}
{\bf P}_{1,1}&{\bf 0}&{\bf 0}\\
{\bf 0}&{\bf P}_{2,2}&{\bf 0}\\
{\bf 0}&{\bf 0}&{\bf P}_{3,3}\\
\end{array}
\right]_{24\times 24},$$
is the matrix of an $(8,16)$-nested interleaver. Each
${\bf P}_{j,j}$ is another permutation, $1\leq j\leq3$, coming from an
interleaver of size $8$. In other words, the interleaver corresponding
to ${\bf P}$ can be constructed by the appending
method with underlying interleavers ${\bf P}_{j,j}$, $1\leq j\leq3$.
\end{Example}
The above example benefited from a set of nested turbo codes.
These turbo codes have tail-bited recursive convolutional codes
as their component codes. Also they used nested interleavers.
We can also simply provide a turbo code based on
an interleaver and two terminated convolutional codes. In this case,
Construction A may be used to obtain a turbo lattice. An example of
a lattice constructed using construction A and turbo code
$\mathcal{TC}$ which uses terminated convolutional codes as its constituent codes
is given next.
\begin{Example}~\label{exterminatedexampleTL}
Let
$${\bf G}_A(x)=\left[
\begin{array}{rr}
1&\frac{1+x^2}{1+x+x^2}
\end{array}
\right]$$
be the generator matrix of a recursive convolutional code.
In this case, we have $K=1$, $N=2$ and $\nu=2$. Let $L=8$; then
we get a turbo code $\mathcal{TC}$ of rate $\frac{LK}{(2\frac{N}{K}-1)(LK+\nu)}=\frac{8}{30}$.
If we use terminated version of these recursive convolutional codes along
with an interleaver of size $8$, a linear block code $\mathcal{TC}[30,8]$
can be obtained. Now consider this turbo code
as a base code of Construction A to induce a turbo lattice.
The minimum distance, coding gain and kissing number of this
turbo lattice is closely related to the minimum distance of
its underlying turbo code. Since the minimum distance of this turbo
code can be increased or decreased by a selection of interleaver, the performance
analysis of this turbo lattice relies on the choice of
its interleaver.
\end{Example}

\section{Parameter Analysis and design Criteria of Turbo Lattice}~\label{ParameterAnalysis}
In this section some
fundamental properties of turbo lattices such as minimum distance,
coding gain and kissing number are studied. These properties give us the possibilities to obtain information
from the underlying turbo codes in order to theoretically check the
efficiency of the constructed turbo lattices.

\subsection{Minimum Distance, Coding Gain and Kissing Number of Turbo Lattices}~\label{AnalysisofTurboLattices}
We look at the turbo lattice $\Lambda_{\mathcal{TC}}$ closer.
The next theorem provides some formulas and an inequality about performance measures
of a turbo lattice $\Lambda_{\mathcal{TC}}$ constructed following Construction D.
\begin{Theorem}~\label{th:propertiesofTL}
 Let $\Lambda_{\mathcal{TC}}$ be a turbo lattice constructed following
Construction D with nested turbo codes
$$\mathcal{TC}=\mathcal{TC}_1\supseteq \mathcal{TC}_2\supseteq\cdots\supseteq \mathcal{TC}_a$$
as its underlying linear block codes with parameters
$[n,k_{\ell},d_{\min}^{(\ell)}]$ and rate $R_{\ell}=\frac{k_{\ell}}{n}$, for $1\leq {\ell}\leq a$. Then
the minimum distance of $\Lambda_{\mathcal{TC}}$ satisfies
\begin{equation}~\label{eq:mindistTL}
 d_{\min}^2(\Lambda_{\mathcal{TC}})=\min_{1\leq {\ell}\leq a}\left\{4,\frac{d_{\min}^{(\ell)}}{4^{{\ell}-1}}\right\}.
\end{equation}
The coding gain is
\begin{equation}~\label{eq:codinggainTL}
 \gamma(\Lambda_{\mathcal{TC}})= 4^{\left(\sum_{{\ell}=1}^aR_{\ell}\right)-1}\min_{1\leq {\ell}\leq a}\left\{4,\frac{d_{\min}^{(\ell)}}{4^{{\ell}-1}}\right\},
\end{equation}
and for the normalized kissing number of $\Lambda_{\mathcal{TC}}$ we have
\begin{equation}~\label{eq:kissnumTL}
 \tau^\ast(\Lambda_{\mathcal{TC}})\leq 2+\sum_{\substack{1\leq {\ell}\leq a\\ d_{\min}^{(\ell)}\leq4^{\ell}}}\frac{2^{d_{\min}^{(\ell)}}}{n}A_{d_{\min}^{(\ell)}},
\end{equation}
where $A_{d_{\min}^{(\ell)}}$ denotes the number of codewords
in $\mathcal{C}_{\ell}$ with minimum weight $d_{\min}^{(\ell)}$.
\end{Theorem}
The proof is given in Appendix A.

\begin{remark}
If the interleaver size $k$ and its relative parameters
$(k_a,\ldots,k_1)$ are increased by a factor of $t$, then the dimension
of the constructed lattice $\Lambda_{\mathcal{TC}}$ increases by the same factor.
As mentioned before, by this modification and due to the interleaver gain and
spectral thining, the minimum distance of the nested turbo codes,
$d_{\min}^{(\ell)}$'s, increase slightly or remain unchanged.
This increase can not be faster than logarithmically with the
code length $n$~\cite{mindist}. Thus, in~(\ref{eq:kissnumTL}), $\frac{2^{d_{\min}^{(\ell)}}}{n}$
decreases. Also the number of minimum weight codewords in these turbo
codes decreases by a factor of $t$.
Hence, the equation~(\ref{eq:kissnumTL})
for the normalized kissing number of $\Lambda_{\mathcal{TC}}$ decreases.
\end{remark}

Now, let us put all the above discussion together.
We can control (increasing of)
coding gain of the constructed turbo lattice $\Lambda_{\mathcal{TC}}$ only
by setting up a good interleaver of size $k$ and adjusting its size.
Furthermore, if one
produces a set of nested turbo codes
$$\mathcal{TC}=\mathcal{TC}_1\supseteq \mathcal{TC}_2\supseteq \cdots \supseteq \mathcal{TC}_a$$
where $d_{\min}^{(\ell)}>\frac{4^{\ell}}{\beta}$ such that $\beta=1$ or $2$, then we get the following bounds
$$d_{\min}(\Lambda_{\mathcal{TC}})\geq\frac{4}{\beta},\quad~\gamma(\Lambda_{\mathcal{TC}})\geq \frac{4^{\sum_{{\ell}=1}^aR_{\ell}}}{\beta},$$
and
$$\tau(\Lambda_{\mathcal{TC}})\leq 2n\quad \mbox{or} \quad \tau^\ast(\Lambda_{\mathcal{TC}})\leq 2.$$

It is obvious that this setting results in (possibly) larger (or at the worst scenario, equivalent) minimum distance,
absolutely better coding gain and (possibly) lower (or at the worst scenario, equivalent) kissing number when compared
with the turbo lattices which come from parallel concatenated of terminated recursive convolutional codes and Construction A.
However, geometrical and layer properties of an $a$ level Construction D turbo lattices
make their decoding algorithm more complex.

According to the discussion described above,
we can take advantage from a wide range of aspects of these
lattices. To be more specific, these turbo lattices are generated by Construction
D using a nested set of block turbo codes. Their underlying codes
are two tail-biting recursive convolutional codes.
Thus, this class provides an appropriate link between two approaches of
block and convolutional codes. The tail-biting method gives us the opportunity
to combine profits of recursive convolutional codes (such as memory) with the
advantages of block codes. It is worth pointing out that the
nested property of turbo codes induces higher coding gain;
see~(\ref{eq:codinggainTL}). Also, excellent performance of parallel
concatenating systematic feed-back convolutional codes imply efficient
turbo lattices with great fundamental parameters.

\subsection{Guidelines to Choose Suitable Parameters}
Since our first priority in designing turbo lattices is to
have high coding gain lattices, selecting appropriate
code length for underlying turbo codes seems crucial.
In addition, guidelines to choose tail-biting convolutional codes that are especially suited for
parallel concatenated schemes are given in~\cite{wiss}. The authors of~\cite{wiss}
also tabulate tail-biting convolutional codes of different rate and length. The minimum distance
of their associated turbo codes are also provided.
We express the importance of parameters like $k_1,\ldots,k_a$ and code length $n$
of underlying turbo codes via a detail example provided below.

Assume that a tail-biting version of
a systematic recursive convolutional code of rate $\frac{2}{3}$ with memory $3$ and
generator matrix
$${\bf G}_1=\left(
\begin{array}{ccc}
1&0&\frac{1+x+x^2+x^3}{1+x^2+x^3}\\
0&1&\frac{1+x+x^2}{1+x^2+x^3}
\end{array}
\right)$$
is used to form a nested turbo code. The resulting turbo code
has rate $R_1=\frac{1}{2}$ and based on~\cite{wiss}, it has minimum distance
$d_{\min}^{(1)}=13$ for block information bits of
length $400$.
Now consider only the first row of the generator matrix
for $\mathcal{TC}_1$. Therefore, the component encoders of $\mathcal{TC}_2$
have generator matrices (after puncturing out the zero bits)
$${\bf G}_2=\left(
\begin{array}{cc}
1&\frac{1+x+x^2+x^3}{1+x^2+x^3}
\end{array}
\right).$$

A block turbo code which uses ${\bf G}_2$ as its constituent codes
has rate $R_2=\frac{1}{3}$ and according to the information in~\cite{wiss},
the minimum distance of this code is $d_{\min}^{(2)}=28$
for information block length of $576$.
For instance suppose that a block of information bits of size $1000$ is used.
Since $\mathcal{TC}_1$ is a rate-$\frac{1}{2}$ block turbo code, the lattice points are in
$\mathbb{R}^{2000}$. Therefore, a square generator matrix of size $2000$
for this turbo lattice ${\bf G}_{\mathcal{TL}}$ can be formed following the approach in Example~\ref{extailbitingexampleTL}.
Hence, ${\bf G}_{\mathcal{TL}}$ is
$$\left[
\begin{array}{rrrr}
\frac{1}{2}{\bf I}_{576}&{\bf 0}&\frac{1}{2}{\bf F}_{1,3}&\frac{1}{2}{\bf P}_{1,1}{\bf F}_{1,3}\\
{\bf 0}&{\bf I}_{324}&{\bf F}_{2,3}&{\bf P}_{2,2}{\bf F}_{2,3}\\
\hline
{\bf 0}&{\bf 0}&2{\bf I}_{500}&{\bf 0}\\
{\bf 0}&{\bf 0}&{\bf 0}&2{\bf I}_{500}
\end{array}
\right],$$
where
$${\bf P}=\left[
\begin{array}{cc}
{\bf P}_{1,1}&{\bf 0}\\
{\bf 0}&{\bf P}_{2,2}
\end{array}
\right]$$
of size $1000\time 1000$ is a $576$-nested interleaver. In other words ${\bf P}$
is an interleaver for $\mathcal{TC}_1$ and ${\bf P}_{1,1}$ is an interleaver
of size $576$ for $\mathcal{TC}_2$. Now the fundamental parameters of
this turbo lattice $\Lambda_{\mathcal{TC}}$ constructed with $2$ levels of Construction D can be found.
Since $d_{\min}^{(1)}=13$ and $d_{\min}^{(2)}=28$, Theorem~\ref{th:propertiesofTL}
implies that
\begin{eqnarray}
d_{\min}^{2}(\Lambda_{\mathcal{TC}})&=&\min_{1\leq \ell\leq2}\left\{4,\frac{d_{\min}^{(1)}}{4^{1-1}},\frac{d_{\min}^{(2)}}{4^{2-1}}\right\}\nonumber\\
&=&\min_{1\leq \ell\leq2}\left\{4,\frac{13}{1},\frac{28}{4}\right\}=4.\nonumber
\end{eqnarray}
and the coding gain of $\Lambda_{\mathcal{TC}}$ satisfies
\begin{eqnarray}
\gamma(\Lambda_{\mathcal{TC}})&=&4^{\sum_{\ell=1}^2R_{\ell}-1}d_{\min}^{2}(\Lambda_{\mathcal{TC}})\nonumber\\
&\geq&4^{\frac{1}{2}+\frac{1}{3}-1}4=4^{\frac{5}{6}},\nonumber
\end{eqnarray}
that is, in decibels, $5~\mbox{dB}$.
Also the kissing number of $\Lambda_{\mathcal{TC}}$ is bounded above by
$$\tau(\Lambda_{\mathcal{TC}})\leq 2n+\sum_{\substack{1\leq {\ell}\leq 2\\ d_{\min}^{(\ell)}\leq4^{\ell}}}2^{d_{\min}^{(\ell)}}A_{d_{\min}^{(\ell)}}.$$
Since $d_{\min}^{(1)}> 4$ and $d_{\min}^{(1)}> 4^2$, the summation in the above inequality disappears
and we get $\tau(\Lambda_{\mathcal{TC}})\leq 4000$ or equivalently $\tau^\ast(\Lambda_{\mathcal{TC}})\leq 2$.

\subsection{Other Possible Design Criteria}
The results in~\cite{forneyspherebound}
provide a general guideline on the choice of code rates
$R_\ell$, $1\leq \ell\leq a$ which is critical
in the construction of any capacity-achieving lattice using Construction D.
Hence a complete different line of studies can be done in order to
change the above design criteria for turbo lattices via
information theoretic tools.

\section{Decoding Algorithm}~\label{DecodingAlgorithm}
There exist many decoding algorithms for finding the closest
point in a lattice~\cite{conway, Viterbo99}.
Similar expressions, algorithms and theorems can be found in~\cite{boundeddistforney}.
In fact, in~\cite{boundeddistforney}, Forney uses a code formula
along with a multi-stage decoding algorithm to solve a CVP for a lattice
based on Construction D.


\subsection{A Multi-Stage Turbo Lattice Decoder}~\label{AMultiStageTurboLatticeDecoder}
In the previous sections we used a set of nested turbo codes
to produce turbo lattice $\Lambda_{\mathcal{TC}}$.
Now our aim is to solve a closest lattice point problem for $\Lambda_{\mathcal{TC}}$.
Assume that a vector ${\bf x}\in\Lambda_{\mathcal{TC}}$ is sent over an unconstrained AWGN channel
with noise variance $\sigma^2$ and a vector ${\bf r}\in\mathbb{R}^n$ is received.
The closest point search algorithms attempt to compute the lattice vector $\tilde{{\bf x}}\in\Lambda_{\mathcal{TC}}$
such that $\|\tilde{{\bf x}}-{\bf r}\|$ is minimized.

The excellent performance of turbo codes is due to the well-known
iterative turbo decoder~\cite{berrou}. One can generalize and investigate
a multi-stage soft decision decoding algorithm~\cite{forneyspherebound} for decoding
lattices constructed based on Construction D. A simple extension to turbo lattices
is presented next.

As it is shown in Section~\ref{BackgroundsonLattices},
every lattice constructed using Construction D benefits from a nice layered code structure.
This building block consists of a set of nested linear block codes which is a set
of nested turbo codes in turbo lattices. The goal is to use $a$, the number of
levels of the construction, serially matching iterative turbo decoding algorithms.
The idea has been brought here from the multi-stage decoding algorithm presented in~\cite{boundeddistforney}.

One can restate~(\ref{eq:codeformulaforD}) as
\begin{equation}~\label{eq:restatecodeformulaforD}
\Lambda_0=2^{a-1}\Lambda_{\mathcal{TC}}=\mathcal{TC}_a+2\mathcal{TC}_{a-1}+\cdots+2^{a-1}\mathcal{TC}_1+2^{a}(\mathbb{Z})^n.
\end{equation}
The above representation of $\Lambda_{\mathcal{TC}}$ states that
every ${\bf x}\in\Lambda_{\mathcal{TC}}$ can be represented by
\begin{equation}~\label{eq:representationofeverylatticepointTC}
\left\{\begin{array}{l}
{\bf x}={\bf x}_1+\frac{1}{2}{\bf x}_2+\cdots+\frac{1}{2^{a-1}}{\bf x}_{a}+{\bf w},~\mbox{or}\\
\\
2^{a-1}{\bf x}={\bf x}_a+2{\bf x}_{a-1}+\cdots+2^{a-1}{\bf x}_1+2^{a-1}{\bf w},
\end{array}\right.
\end{equation}
where ${\bf x}_{\ell}\in \mathcal{TC}_{\ell}$ and ${\bf w}\in2(\mathbb{Z})^n$, $1\leq {\ell}\leq a$.

Any soft-input soft-output (SISO) or soft-input hard-output (SIHO) decoding algorithm for the turbo code $\mathcal{TC}_\ell$ may
be used as a decoding algorithm for $\Lambda_{\ell}=\mathcal{TC}_\ell+2\mathbb{Z}^n$, as follows.
Given any ${\bf r}_{\ell}$, let us denote the closest even and odd
integers to each coordinate
$r_j$ of ${\bf r}_{\ell}$ by $e_j$ and $o_j$ respectively, $1\leq j\leq n$. Then one can compute
$t_j=2(\pm\frac{e_j+o_j}{2}\mp r_j)$
(where the upper signs are taken if $e_j<o_j$
and the lower ones if $e_j>o_j$)
and consider $t_j$ as the ``metric" for $0$ and $1$, respectively.
Then the vectors ${\bf t}_{\ell}=(t_1,\ldots,t_n)$ (as the confidence vector) and
${\bf s}_{\ell}={\bf r}_{\ell}\pmod{2}$ (as the received vector) are passed
to a SISO (or SIHO) decoder for $\mathcal{TC}_\ell$.
The decoded turbo codeword is then mapped back to $e_j$ or $o_j$, at the $j$--th coordinate, depending on whether the decoded
codeword is $0$ or $1$ in that coordinate.

The above algorithm is for $\Lambda_\ell$.
A general scheme for this multi-stage turbo lattice decoder
can be shown by the following simple pseudo-code.
A similar algorithm for a Construction D Barnes-Wall lattices
can be found in~\cite{Harshan13}.

\noindent
{\bf Decoding Algorithm for Turbo Lattices}\\
{\bf Input:} {\bf r} an $n$-dimensional vector in $\mathbb{R}^n$.\\
{\bf Output:} a closest vector $\tilde{{\bf x}}$ to ${\bf r}$ in $\Lambda_{TC}$.
\begin{itemize}
 \item{}{\bf Step 1)}  Put ${\bf r}_a=2^{a-1}{\bf r}$.
 \item{}{\bf Step 2)}\\
 {\bf for} $\ell=a$ {\bf downto} $1$ {\bf do}
 \begin{itemize}
 \item{} Decode ${\bf r}_{\ell}$ to the closest point ${\bf x}_{\ell}\in\Lambda_{\ell}=C_{\ell}+2(\mathbb{Z})^n$.
 \item{} Compute ${\bf r}_{\ell-1}=\frac{{\bf r}_{\ell}-{\bf x}_{\ell}}{2}$.
 \end{itemize}
\item{}{\bf Step 3)} Evaluate $\tilde{{\bf x}}=\frac{{\bf x}_a+2{\bf x}_{a-1}+\cdots+2^{a-1}{\bf x}_1+2^{a-1}{\bf w}}{2^{a-1}}$.
\end{itemize}

The next theorem shows that the above algorithm can find the closest lattice
point of $\Lambda_{\mathcal{TC}}$ to the received vector ${\bf r}$ when the points of
$\Lambda_{\mathcal{TC}}$ are sent over an unconstrained AWGN channel with noise variance $\sigma^2$.
There is a similar theorem and proof in~\cite{boundeddistforney}, however for the sake of
completeness we give them both in the following.
\begin{Theorem}~\label{th:truedecoder}
 Given an $n$-tuple ${\bf r}$, if there is a point $\tilde{{\bf x}}$
in $\Lambda_{\mathcal{TC}}$ such that $\|{\bf r}-\tilde{{\bf x}}\|^2< d_{\min}^2(\Lambda_{\mathcal{TC}})/4$,
then the algorithm decodes ${\bf r}$ to $\tilde{{\bf x}}$.
\end{Theorem}
The proof is given in Appendix A.
In the next subsection we analyze the decoding complexity of the proposed algorithm.

\subsection{Decoding Complexity}~\label{DecodingComplexity}
Since the operations for computing
the nearest odd and even integer numbers close to the components of a received vector
${\bf r}$ are negligible, the decoding complexity of a lattice $\Lambda_{\ell}=\mathcal{C}_{\ell}+2(\mathbb{Z})^n$ constructed
using Construction A is equal to the complexity of decoding the turbo code
$\mathcal{C}_{\ell}$ via an iterative turbo decoder.
As shown before, a turbo lattice decoder uses exactly $a$ subsequent and successive
turbo decoder algorithms for $\Lambda_{\ell}$, $1\leq {\ell}\leq a$. Thus
the overall decoding complexity of the proposed turbo lattice decoding algorithm can not
exceed $a$ times the decoding complexity of an iterative turbo decoder.

\subsection{Other Possible Decoding Methods}~\label{OtherPossibleDecodingMethods}
\begin{figure*}[t!]%
  \begin{center}%
  \vspace{1.5cm}
\includegraphics[width=10cm]{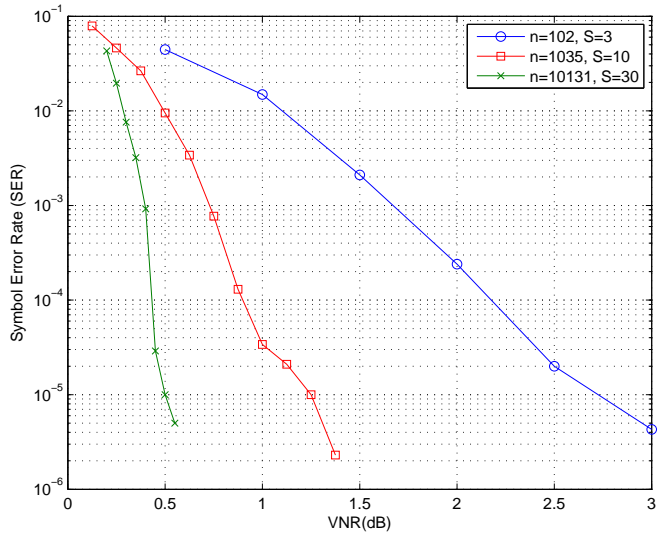}~\caption{Comparison graph for various lengths of a turbo lattice.}~\label{fig:comarison}
  \end{center}
\end{figure*}

The multi-stage decoding algorithm may not be the best choice here.
Some other options are listed in the following paragraphs.

First, for low dimensional lattices a universal lattice code
decoder~\cite{Viterbo99} can be employed to decode
turbo lattices. In that case, one can carve good lattice
constellations from turbo lattices by choosing appropriate
shaping regions~\cite{Boutros96}.

Second, it is well-known that turbo codes can be considered as a class of graph-based codes
on iterative decoding~\cite{recent}.
Also Tanner graph realization of lattices are introduced
in~\cite{Banihashemi01}. In fact in a multi-stage decoding,
after a ``coarse" code is decoded, it is frozen and the decision
is passed to a ``fine" code. In other words,
there is no iterative decoding across layers.
If the nested underlying turbo codes
of a turbo lattice are expressed as a Tanner-graph model as in~\cite{recent}
or the turbo lattice itself is presented by a Tanner-graph as in~\cite{Banihashemi01},
then it seems feasible to do iterative decoding across layers,
which may potentially increase the performance.
However we should again be careful about short cycles in the corresponding
Tanner-graph of turbo lattices as well as
cycle-free lattices~\cite{sakzad10}.

Third, lattice basis reduction algorithms and the faster version
of that which works in complex plane~\cite{Ling09} can also be employed
to find closest lattice point to a received vector.

\section{Simulation Results for Performance of Turbo Lattices}~\label{SimulationResultsPerformanceofTL}
All simulation results presented
are achieved using an AWGN channel, systematic recursive
convolutional codes in the parallel
concatenated scheme, and iterative turbo lattice
decoder all discussed earlier. Indeed, we investigate turbo lattice designed using two
identical terminated convolutional codes with generator matrix
${\bf G}_A(x)$. Turbo lattices of different
lengths are examined. Furthermore, the performance of these
turbo lattices are evaluated using BCJR algorithms~\cite{costello} as constituent
decoders for the iterative turbo decoder.
Moreover, $S$-random interleavers of sizes $2^5$, $7^3$ and $15^3$ such that
$S$ equals to $3$, $10$ and $30$ have been used, respectively. These results in turbo lattices of dimensions
$(2^5+2)*3=102$, $(7^3+2)*3=1035$ and $(15^3+2)*3=10131$. The number of
iterations for the iterative turbo decoder is fixed. It is equal to ten in all cases.
Fig. 1 shows a comparison between turbo lattices formed with
turbo codes of different lengths. These turbo lattices
achieve a symbol error rate (SER) of $10^{-5}$ at an $\alpha^2=2.75~\mbox{dB}$ for size $102$,
an $\alpha^2=1.25~\mbox{dB}$ for frame length $1035$.
Also an SER of $10^{-5}$ is attained at an $\alpha^2=.5~\mbox{dB}$ for size $10131$.

In the following we compare these results for turbo lattices with other
newly introduced latices including LDPC lattices~\cite{sadeghi}
and LDLC lattices~\cite{LDLC}. The comparison is presented in Table~\ref{tableofcomparison}.

\begin{table}[h!]
\begin{center}
\begin{tabular}{c|c|c|c}
\hline\hline
Lattice & $n$ & Error Probability & Distance from Capacity  \\
\hline\hline
LDPC Lattice & $2000$ & NEP$=10^{-5}$&$2.8$~dB\\
\hline
LDLC Lattice & $1000$ & SER$=10^{-5}$&$1.5$~dB\\
\hline
Turbo Lattice & $1035$ & SER$=10^{-5}$&$1.25$~dB\\
\hline\hline
\end{tabular}
\end{center}\caption{A comparison between well-known and newly introduced lattices.}~\label{tableofcomparison}
\end{table}
In Fig. 3 and for turbo lattices of sizes $n=102,~1035,~10131$, at SER of $10^{-5}$,
we achieve $\alpha^2=2.75,~1.25$ and $.5$ dB away from capacity
while for $n=100,~1000,~10000,~100000$, LDLC lattices~\cite{LDLC} can work as close
as $3.7,~1.5,~0.8$ and $0.6$ dB from capacity, respectively.
Thus, we have an excellent performance of turbo lattices when compared with other lattices.

\section{Conclusion and Further Research Topics}~\label{Conclusion}
The concept of turbo lattices is
established using Construction D method for lattices along
with a set of newly introduced nested turbo codes.
To this end, tail-biting and
terminated convolutional codes are concatenated in parallel.
This parallel concatenation to induce turbo codes was
supported with nested $S$-random interleavers.
This gives us the possibility to combine the characteristics of
convolutional codes and block codes to produce good turbo lattices.
The fundamental parameters of turbo lattices for investigating
the error performance are provided. This includes minimum distance, coding gain,
kissing number and an upper bound on the probability of error.
Finally, our experimental results show excellent performances for
turbo lattices as expected by the theoretical results.
More precisely, for example at SER of $10^{-5}$ and for $n=10131$
we can work as close as $\alpha^2=0.5$ dB from capacity.

Analyzing other factors
and parameters of $b$-branches turbo lattices such as sphere packing, covering and
quantization problem is also of great interest.
Another interesting research problem is to find the error performance
of turbo lattices designed by other types of interleavers including
deterministic interleavers~\cite{allertonversion, recent}.

Since the performance of turbo lattices depends on
the performance of their underlying codes, then search
for other well-behaved turbo-like codes would be interesting.

%

\appendix[A:Proofs]
{\em Theorem~\ref{th:constructionDmindistkiss}:}
\begin{itemize}
\item{}
Let ${\bf c}^{\ell}$ be a codeword with minimum weight in $\mathcal{C}_{\ell}$ for $1\leq\ell\leq a$.
There exist $\beta_j^{({\ell})}\in\{0,1\}$ such
that ${\bf c}^{\ell}=\sum_{j=1}^{k_{\ell}}\beta_j^{({\ell})}{\bf c}_j$.
Since $\frac{1}{2^{{\ell}-1}}{\bf c}^{\ell}$ is in the form
of~(\ref{eq:formofconstructionD}), it belongs to $\Lambda$.
Thus, we have
$$d_{\min}(\Lambda)\leq\left\|\frac{1}{2^{{\ell}-1}}{\bf c}^{\ell}\right\|^{1/2}=\frac{1}{2^{{\ell}-1}}\sqrt{d_{\min}^{(\ell)}}.$$
This means that
$d_{\min}(\Lambda)\leq\min_{1\leq {\ell}\leq a}\left\{2,\frac{1}{2^{{\ell}-1}}\sqrt{d_{\min}^{(\ell)}}\right\}$.
On the other hand the number $2$ in this formula happens when $\beta_{j}^{(\ell)}=0$
for all $j$ and $\ell$ and ${\bf z}=(0,\ldots,0,2,0,\ldots,0)$.
Now, we set $L_0=(2\mathbb{Z})^n$ and
$$L_{\ell}=\left\{{\bf x}+\sum_{j=1}^{k_{\ell}}\beta_j^{({\ell})}\frac{{\bf c}_j}{2^{{\ell}-1}}\right\}$$
where $\beta_j^{({\ell})}=0$ or $1$ and ${\bf x}\in L_{{\ell}-1}$. Hence,
$L_a=\Lambda$. It is easy to check that $L_{\ell}$
is a lattice. Let $0\neq{\bf v}$ be a vector of level ${\ell}$. If
${\ell}=0$ then $\|{\bf v}\|=4$. If ${\ell}>0$ then
according to the definition of $L_{\ell}$ we can write ${\bf v}={\bf x}+{\bf y}$
where ${\bf x}\in L_{{\ell}-1}$ and ${\bf y}\in L_{\ell}$. The vector ${\bf y}$ has at least
$d_{\min}^{(\ell)}$ components since it is in level ${\ell}$. It means that the norm of the vector
${\bf y}$ is at least $\frac{d_{\min}^{(\ell)}}{4^{{\ell}-1}}$. So ${\bf y}$ has at least
$d_{\min}^\ell$ components equaling to $(1/2)^{\ell-1}$. Note also that every
component in ${\bf x}$ is a multiple of $(1/2)^{\ell-2}$. Hence ${\bf v}$ has at least
$d_{\min}^\ell$ components whose absolute values are no less than $(1/2)^{\ell-1}$.
It follows that the norm of the vector
${\bf v}$ is at least $\frac{d_{\min}^{(\ell)}}{4^{{\ell}-1}}$
Thus, $d_{\min}(\Lambda)\geq\min_{1\leq {\ell}\leq a}\left\{2,\frac{1}{2^{{\ell}-1}}\sqrt{d_{\min}^{(\ell)}}\right\}$.
\item{}
The only points in $\Lambda$ that achieve $d_{\min}(\Lambda)$ are the
$2n$ points $\pm2{\bf e_i}$ for $1\leq i\leq n$ where ${\bf e_i}$
is the $i$--th unit vector plus the points which are in
$\mathcal{C}_{\ell}$'s satisfying~(\ref{eq:mindistD}) for $1\leq {\ell}\leq a$.
In other words when we have
$d_{\min}^2(\Lambda)=\frac{d_{\min}^{(\ell)}}{4^{\ell-1}}$
for some $1\leq {\ell}\leq a$,
then $4\geq\frac{d_{\min}^{(\ell)}}{4^{\ell-1}}$ and the codewords
with minimum weight in $\mathcal{C}_{\ell}$ are the candidates
to produce spheres which can touch the sphere with center $0$ and radius $\frac{d_{\min}(\Lambda)}{2}$.
Therefore, these points must be in $\mathcal{C}_{\ell}$
with weight $d_{\min}^{(\ell)}$ such that $d_{\min}^{(\ell)}\leq4^{{\ell}}$.
It means that the kissing number of
$\Lambda$ is upper bounded by
\begin{equation}~\label{eq:ub}
2n+\sum_{\substack{1\leq {\ell}\leq a\\ d_{\min}^{(\ell)}\leq4^{\ell}}}2^{d_{\min}^{(\ell)}}A_{d_{\min}^{(\ell)}}.
\end{equation}
The coefficient $2^{d_{\min}^{(\ell)}}$ appears since the nonzero
entries of each lattice vector of $\Lambda$ can be positive or negative.
We note that if $d_{\min}^{(\ell)}>4^{{\ell}}$,
then the right hand side of~(\ref{eq:mindistD}) is equal to $2$ and the summations in~(\ref{eq:ubkiss})
and~(\ref{eq:ub}) disappear.
\end{itemize}
{\em Theorem~\ref{th:converCCsys}:}
The vector
$$(0,\ldots,0,1,0,\ldots,0,f_{i,K+1}(x),\ldots,f_{i,N}(x))$$
is in the code if and only if there exists a polynomial $t_i(x)$
such that
$$t_i(x)\left(0,\ldots,0,r_i(x),0,\ldots,0,q_{i,K+1}(x),\ldots,q_{i,N}(x)\right)\equiv$$
\vspace{-.5cm}
$$(0,\ldots,0,1,\ldots,0,f_{i,K+1}(x),\ldots,f_{i,N}(x))\!\!\!\!\pmod{x^L-1}.$$
This happens if and only if every $r_i(x)$ has an inverse $\mbox{mod}~(x^L-1)$
and $q_{i,j}(x)\equiv f_{i,j}(x)r_i(x)\pmod{x^L-1}$ for
$1\leq i\leq K$ and $K+1\leq j\leq N$ because
$r_i^{-1}(x)=t_i(x)$ for $1\leq i\leq K$.

{\em Proposition~\ref{relation}:}
Since the characteristic polynomial of the matrix ${\bf A}$ is $r(x)$, we conclude that $p({\bf A})=0$.
We have that $\gcd(r(x),x^L-1)=1$ if and only if there exist two polynomials $s(x)$ and $t(x)$ such that
$s(x)r(x)+t(x)(x^L-1)=1$. Put $x={\bf A}$, we get $t({\bf A})({\bf A}^L-{\bf I}_m)={\bf I}_m$. Hence,
we have $\det\left({\bf A}^L-{\bf I}_m\right)\neq0$ if and only if $\gcd(r(x),x^L-1)=1$.

{\em Theorem~\ref{th:unchengedrate}:}
Let $R_{\ell}^0=\frac{k_{\ell}^0}{n_{\ell}^0}$ denote the rate of
the ${\ell}$--th component of the nested turbo codes when we scale $L$
and $k_{\ell}$'s by a factor of $t$. Then, we have
$$R_{\ell}^0=\frac{k_{\ell}^0}{n_{\ell}^0}=\frac{tk_{\ell}}{(2(Lt)N-(Lt)K)}=R_{\ell}.$$
We observe that in this case the interleaver size is $k^0=tk=(Lt)K$
and the interleaver $\Pi$ is a $(tk_a,\ldots,tk_1)$-nested interleaver.
Also this is true for the actual rate of our nested turbo codes.
Suppose $n_{\ell}^0, k_{\ell}^0$ and $R_{\ell}^0$ be as above, then
\begin{eqnarray}
   R_{\ell}^0&=&\frac{k_{\ell}^0}{n_{\ell}^0-k^0+k_{\ell}^0}=\frac{tk_{\ell}}{(2(Lt)N-(Lt)K)-tk+tk_{\ell}}\nonumber\\
   &=& \frac{k_{\ell}}{n-k+k_{\ell}}=R_{\ell}.\nonumber
\end{eqnarray}

{\em Theorem~\ref{th:propertiesofTL}:}
By using Theorem~\ref{th:constructionDmindistkiss}
and the paragraph above that, we easily get~(\ref{eq:mindistTL}) and~(\ref{eq:kissnumTL}).
For the coding gain we have
\begin{eqnarray}
   \gamma(\Lambda_{\mathcal{TC}})&=&\frac{d^2_{\min}(\Lambda_{\mathcal{TC}})}{\det(\Lambda_{\mathcal{TC}})^{2/n}}=\frac{\min_{1\leq {\ell}\leq a}\left\{4,\frac{d_{\min}^{(\ell)}}{4^{{\ell}-1}}\right\}}{\left(2^{n-\sum_{{\ell}=1}^ak_{\ell}}\right)^{2/n}}\nonumber\\
   &=& 4^{\left(\sum_{{\ell}=1}^aR_{\ell}\right)-1}\min_{1\leq {\ell}\leq a}\left\{4,\frac{d_{\min}^{(\ell)}}{4^{{\ell}-1}}\right\}.\nonumber
\end{eqnarray}

{\em Theorem~\ref{th:truedecoder}:}
Based on Equations~(\ref{eq:mindistD}) and~(\ref{eq:restatecodeformulaforD}),
and the fact that $d_{\min}^2(2^{a-1}\Lambda_{\mathcal{TC}})=4^{a-1}d^2_{\min}(\Lambda_{\mathcal{TC}})$,
we get
$$d_{\min}^2(\Lambda_0)=4^{a-1}\min_{1\leq {\ell}\leq a}\left\{4,\frac{d_{\min}^{(\ell)}}{4^{{\ell}-1}}\right\}.$$
In addition, based on the geometric uniformity of lattices, it suffices to
consider $\tilde{{\bf x}}=\tilde{{\bf y}}={\bf 0}$. Therefore, we have to prove that if
$\|{\bf r}_a\|^2<d_{\min}^2(\Lambda_0)/4$, then ${\bf r}_a$ will be decoded to
${\bf 0}$. There is an error
in step ${\ell}$ if ${\bf x}_{{\ell}}\neq0$, $1\leq {\ell}\leq a$. Assume that there have been no errors
at the former steps $1\leq {\ell}'<{\ell}$. Then in the ${\ell}$--th step we have ${\bf r}_{a-{\ell}+1}=\frac{{\bf r}_a}{2^{\ell-1}}$.
Hence,
$$\|{\bf r}_{a-{\ell}+1}\|^2=\frac{1}{4^{{\ell}-1}}\|{\bf r}_a\|^2\leq
\frac{d_{\min}^2(\Lambda_0)}{4^{\ell}}$$
$$\leq 4^{a-1}\frac{d_{\min}^{(a-{\ell}+1)}}{(4^{a-{\ell}})(4^{\ell})}=\frac{d_{\min}^{(a-{\ell}+1)}}{4}.$$
This means that the largest sphere that can be inscribed in the Voronoi region
associated with the lattice vector ${\bf 0}$ has radius
$\frac{d_{\min}^{(a-{\ell}+1)}}{4}$.
Since
$\|{\bf r}_{a-{\ell}+1}\|^2<\frac{d_{\min}^{(a-{\ell}+1)}}{4}$,
the point ${\bf r}_{a-{\ell}+1}$ is in the sphere of
lattice $\Lambda_{a-{\ell}+1}=\mathcal{C}_{a-{\ell}+1}+2(\mathbb{Z})^n$, and thus no error can
occur at step ${\ell}$.

\ifCLASSOPTIONcaptionsoff
  \newpage
\fi

\end{document}